\documentstyle[12pt,twoside,fleqn,espcrc1]{article}

\newcommand{\AmS}{{\protect\the\textfont2
  A\kern-.1667em\lower.5ex\hbox{M}\kern-.125emS}}

\title{ABSSM --- a new field--theoretic method of describing 
scattering/ production processes of composite particles by asymptotic 
in--/ outgoing Bethe--Salpeter states}

\author{F. Kleefeld\address{Centro de F\'{\i}sica das Interac\c{c}\~{o}es Fundamentais (CFIF), Instituto Superior T\'{e}cnico, Edif\'{\i}cio Ci\^{e}ncia, Piso 3, Av. Rovisco Pais, P-1049-001 LISBOA, Portugal}%
        \thanks{This work has been supported by the {\em Funda\c{c}\~{a}o para a Ci\^{e}ncia e a Tecnologia} (FCT) of the {\em Minist\'{e}rio da Ci\^{e}ncia e da Tecnologia} of Portugal under Grant No.\ PRAXIS XXI/BPD/20186/99.}}

\newcommand{\oabs}[1]{|\vec{#1} \makebox[0.1em]{} |}

\newcommand{\slsh}[1]{\, {\not {\! #1}}}

\newcommand{\beq}{\begin{equation}}
\newcommand{\eeq}{\end{equation}}
\newcommand{\beqa}{\begin{eqnarray}}
\newcommand{\eeqa}{\end{eqnarray}}

\begin{document}

\maketitle


The relativistic field theoretic treatment of asymptotic composite states in many particle scattering and production reactions or electromagnetic current distributions has a long tradition (see e.g.\ R.\ Haag, Phys.\ Rev. 112 (1958) 669), yet full application of the formalism even to simple problems gets quite cumbersome. The presented new relativistic field--theoretic formalism called {\em ``Asymptotic Bethe--Salpeter State Method'' (ABSSM)} tries to interpolate between the inclusion of as many present theoretical requirements as possible and the technical and computational capabilities available. 

For a reaction $1+2\rightarrow 1^\prime+\ldots+n^\prime$ the T-matrix $T_{fi}$ is obtained by ($\alpha_i$ denotes further internal quantum numbers, $S_{int}:=\int d^4x \; {\cal L}_{int} (x)$ the interaction part of the action)\footnote{The double left bracket $\ll \ldots |$ reminds the careful reader, that out--states are not obtained from in--states $|\ldots >$ just by a simple Hermitian conjugation.}:
\begin{equation} (2\pi)^4\,\delta^4(P_f-P_i)\;i\, T_{fi} \; = \; 
 \ll p_{1^\prime},\alpha_{1^\prime}; \ldots ; p_{n^\prime},\alpha_{n^\prime} | \,T [e^{\displaystyle \,i\!:S_{int}:} - 1 ] \,| p_{1},\alpha_{1}; p_{2},\alpha_{2} >
\end{equation}
while a typical Breit-frame momentum space current distribution for the calculation of electromagnetic formfactors is given in covariant gauge by ($q:=p^{\,\prime\,\mu}-p^{\,\mu} = (0,\vec{q}\,)$):
\begin{eqnarray}
\lefteqn{j^{\,\mu} (\vec{q}\,;\alpha^{\,\prime},\alpha) 
\; =} \nonumber \\
 & = &
\frac{1}{2\,\omega (|\vec{p}\,|)} \;(\,g^{\,\mu\nu} \, \Box_{\,x}  \,-\, (1\,-\,\zeta )\; \partial_x^{\,\mu} \,\partial_x^{\,\nu}  \,) 
\ll p^{\,\prime},\alpha^{\,\prime}\;|T \left[  \,  A_{\,\nu} (x) \; e^{\displaystyle\,i:S_{int}:}  \,\right] |\,p ,\alpha >\,\Bigg|_{x=0} 
\end{eqnarray}
Assume now, that at least one of the in-- or out--states in (1) and the in-- and out--state in (2) is a composite system denoted here by $|P,B>$, which fulfils the same 
relativistic onshell--normalization condition as the underlying ``elementary'' one--particle states, i.e.:
\begin{equation}
 \ll P^\prime,B^\prime|P,B> = (2\pi)^3 \; 2\, \omega_B(|\vec{P}\,|) \, \delta^{\,3} (\vec{P}^{\,\prime} - \vec{P}\,)\;\delta_{B^\prime B}
\end{equation}
The composite state vector can be decomposed into \underline{\em free} one--particle Fock--states, i.e.:
\begin{eqnarray}
\lefteqn{|P,B> \; = \; |0>\ll 0|P,B>  +  \sum\limits_{\alpha_1} \int \frac{d^3p_1}{(2\pi)^3
\, 2\, \omega_1 (|\vec{p}_1|)} \;
|\vec{p}_1,\alpha_1>\ll \vec{p}_1,\alpha_1|P,B> +} 
 \nonumber \\
 & + & \sum\limits_{\alpha_1,\alpha_2}\!\int\!\!\frac{d^3p_1}{(2\pi)^3
\, 2\, \omega_1 (|\vec{p}_1|)}
\frac{d^3p_2}{(2\pi)^3
\, 2\, \omega_2 (|\vec{p}_2|)} \;
|\vec{p}_1,\alpha_1;\vec{p}_2,\alpha_2>
\ll \vec{p}_1,\alpha_1;\vec{p}_2,\alpha_2|P,B> + \ldots
\end{eqnarray} 
The Fourier--expansion of free Bosonic and Fermionic field operators $\phi (x)$ and $\psi (x)$:
\begin{eqnarray}
 \phi (x) & = & \sum\limits_{s,t} \int \! 
\frac{d^3p}{(2\pi )^3 2 \, \omega (\oabs{p})} 
\; 
[ \;
e^{-\, i px} \, \chi \,(\vec{p},s,t) \; a\,(\vec{p},s,t) +
e^{i px} \, \chi^+ (\vec{p},s,t) \; a^+(\vec{p},s,t) 
\; ] \nonumber \\
\psi (x) & = & \sum\limits_{s,t} \int \! 
\frac{d^3p}{(2\pi )^3 2 \, \omega (\oabs{p})} 
\; 
[ \;
e^{-\, i px} \, u \,(\vec{p},s,t) \; b\,(\vec{p},s,t) + 
e^{i px} \, v \,(\vec{p},s,t) \; d^{\, +}(\vec{p},s,t) 
\; ] \nonumber \\
 & & \nonumber \\
 \lefteqn{( \bar{u}\, (\vec{p} , s) \; u \, (\vec{p} , s^\prime ) = 
2\, m \;\delta_{ss^\prime} \; , \quad\bar{v} \, (\vec{p} , s) \; v \, (\vec{p} , s^\prime )  =  - \,
2\, m \;\delta_{ss^\prime}
)}   
\end{eqnarray}
($s,t$ are spin-- and isospin--projection quantum numbers) can easily inverted by:
\begin{eqnarray}
a \, (\vec{p},s,\ldots) & = & (2\pi )^3 \, 2\, \omega \, (\oabs{p}) \; \cdot  \nonumber \\ 
 & \cdot & \chi^+ \, (\vec{p}, s,\ldots) \;\; \int \frac{d^3x}{(2\,\pi)^3} \;\;
\frac{1}{2} \;
\left( \phi \, (x) + \frac{i \, \dot{\phi} \, (x)}{\omega \, (\oabs{p})} \;  \right)
\Bigg|_{x^0=0} \;\;
e^{\displaystyle\;-\,i\,\vec{p}\cdot \vec{x}}
\nonumber \\ 
 & & \nonumber \\
b \, (\vec{p},s,\ldots) & = & \quad \frac{(2\pi )^3 \, 2\, \omega \, (\oabs{p})}{2m} \;\; \bar{u} \, (\vec{p},s,\ldots) \; \int \frac{d^3x}{(2\,\pi)^3} \;\;\psi (x)\,\Big|_{x^0=0} \;\;\,
e^{\displaystyle\;-\,i\,\vec{p}\cdot \vec{x}}
\nonumber \\
 & & \nonumber \\
d^+ \, (\vec{p},s,\ldots) & = & - \, \frac{(2\pi )^3 \, 2\, \omega \,(\oabs{p})}{2m} \;\;  \bar{v} \, (\vec{p},s,\ldots) \; \int \frac{d^3x}{(2\,\pi)^3} \;\;\psi (x)\,\Big|_{x^0=0} \;\;\,
e^{\displaystyle\;+\,i\,\vec{p}\cdot \vec{x}} 
\end{eqnarray}
These identities are now applied to the matrix elements $\ll \vec{p}_1,s_1;\ldots ;\vec{p}_n,s_n|P,B>$ in (4):
\begin{eqnarray}
\ll \vec{p}_1,s_1;\ldots ;\vec{p}_n,s_n|P,B> & \stackrel{!}{=} & \left( \prod\limits^n_{i=1} \;
\frac{(2\pi )^3 \, 2\, \omega_i \, (|\vec{p}_i|)}{2\,m_{\,i}} \;\; \bar{u}^{(i)}  (\vec{p}_i,s_i) \; \int \frac{d^3x_i}{(2\,\pi)^3} \; 
e^{\displaystyle\;-i\,\vec{p}_i \cdot \vec{x}_i }
\right) \; \cdot \nonumber \\
 & \cdot & \ll 0|\; \psi^{(n)} (x_n) \, \ldots \, \psi^{(1)} (x_1) \, |P,B> 
\Big|_{x_1^0,\ldots,\,x^0_n\rightarrow 0}
\end{eqnarray}
The crucial and {\em only}\ two assumptions of the presented
new ABSSM are the following: \\[1mm]
(a) \parbox[t]{15cm}{Inverse application of the {\em LSZ-reduction technique} is possible and yields $\psi (x) \rightarrow  Z^{-1/2}_\psi \; \psi_H(x)$ ($\psi_H(x) = $  Heisenberg--operator, $Z_\psi = $ renormalization constant).} \\[1mm]
(b) \parbox[t]{15cm}{ Due to {\em causality} the {\em composite state $|P,B>$ projects 
on time--ordered products} of field--operators.}\\[1mm] 
Using these assumptions the ABSSM allows the following replacement in (7):
\begin{eqnarray}
\lefteqn{\ll 0|\; \psi^{(n)} (x_n) \, \ldots \, \psi^{(1)} (x_1) \, |P,B>  
\Big|_{x_1^0,\ldots,\,x^0_n \rightarrow 0} \;\;{\stackrel{!}{\equiv}} } \nonumber \\
 & {\stackrel{!}{\equiv}} & Z^{-n/2}_\psi
\ll 0|\,T [ \psi_H^{(n)} (x_n) \, \ldots \, \psi_H^{(1)} (x_1) ] \, |P,B>  
\Big|_{x^0_1, \ldots,\, x^0_n \rightarrow 0}  
\end{eqnarray}
An analogue expression holds for Bosons. In the following I will skip for simplicity the indices ``$H$'' and set the renormalization constants to one\footnote{Future investigations of course will have to illuminate the detailed role of the renormalization constants!}. For a Fermionic composite system the Fock--expansion (4) of the composite state vector within the ABSSM yields in configuration space (A similar expansion of purely Bosonic and mixed Fermion--Boson composite systems is straight forward!):

\begin{eqnarray} \lefteqn{|P,B>\quad = \quad |0>\ll 0|P,B> \; +} \nonumber \\
 & + & 
\sum\limits_{s_1,\ldots} 
\int \frac{d^3p_1}{(2\pi )^3 \, 2 \,m_1} \int d^3x_1 
\; \exp[- i\,\vec{p}_1\cdot \vec{x}_1] \nonumber \\
& & \Big\{ \;
b^+ \, (\vec{p}_1,s_1,\ldots) |0> \; 
\bar{u}^{(1)} (\vec{p}_1,s_1,\ldots) \;
\ll 0| \; \psi^{(1)}  (x_1) \, |P,B>  \nonumber \\
& & - \;
d^+ \, (\vec{p}_1,s_1,\ldots) |0> \; 
\ll 0| \; \bar{\psi}^{(1)}  (x_1) \, |P,B>  
v^{(1)} (\vec{p}_1,s_1,\ldots) \; \Big\} \Big|_{x^0_1\rightarrow 0}
\; + \nonumber \\
 & + & 
\sum\limits_{s_1,\ldots} \sum\limits_{s_2,\ldots}
\int 
\frac{d^3p_1}{(2\pi )^3 \, 2 \,m_1} 
\frac{d^3p_2}{(2\pi )^3 \, 2 \,m_2}
\int d^3x_1 \; d^3x_2 \;\;
\exp[- i\,\vec{p}_1\cdot \vec{x}_1- i\,\vec{p}_2\cdot \vec{x}_2] 
\nonumber \\
& & \Big\{ \;
b^+ \, (\vec{p}_1,s_1,\ldots) \; b^+ \, (\vec{p}_2,s_2,\ldots) |0>  
\nonumber \\ 
 & & \; \;\;\;
\bar{u}^{(1)} (\vec{p}_1,s_1,\ldots) \;
\bar{u}^{(2)} (\vec{p}_2,s_2,\ldots) \;
\ll 0| \, T \, [ \, \psi^{(2)}  (x_2) \; \psi^{(1)}  (x_1) \, ] \, |P,B>  \nonumber \\
 & & + \;
d^+ \, (\vec{p}_1,s_1,\ldots) \; d^+ \, (\vec{p}_2,s_2,\ldots) |0>  
\nonumber \\ 
 & & \; \;\;\;
\ll 0| \, T \, [ \, \bar{\psi}^{(2)}  (x_2) \; \bar{\psi}^{(1)}  (x_1) \, ] \, |P,B>  
v^{(1)} (\vec{p}_1,s_1,\ldots) \;
v^{(2)} (\vec{p}_2,s_2,\ldots) \nonumber \\
 & & - \;
b^+ \, (\vec{p}_1,s_1,\ldots) \; d^+ \, (\vec{p}_2,s_2,\ldots) |0>  
\nonumber \\ 
 & & \; \;\;\;
\bar{u}^{(1)} (\vec{p}_1,s_1,\ldots) 
\ll 0| \, T \, [ \,  \bar{\psi}^{(2)}  (x_2) \; \psi^{(1)}  (x_1) \, ] \, |P,B>  
 v^{(2)} (\vec{p}_2,s_2,\ldots)  \nonumber \\
 & & - \;
d^+ \, (\vec{p}_1,s_1,\ldots) \; b^+ \, (\vec{p}_2,s_2,\ldots) |0> \; 
\nonumber \\ 
 & & \; \;\;\;
\bar{u}^{(2)} (\vec{p}_2,s_2,\ldots) 
\ll 0| \, T \, [ \,\psi^{(2)}  (x_2) \; \bar{\psi}^{(1)}  (x_1) \, ] \, |P,B>   
 v^{(1)} (\vec{p}_1,s_1,\ldots) \Big\} \Big|_{x^0_1,\,x^0_2 \rightarrow 0} + \ldots 
\end{eqnarray}
After introduction of Jacobi--coordinates --- in the two particle sector they are defined by $X := \eta_1 \, x_1 + \eta_2 \, x_2$, $x := x_1 - \, x_2$, $P := p_1 + p_2$, $q := \eta_2 \, p_1 - \, \eta_1 \, p_2$ with $\eta_1+\eta_2=1$ --- the appearing Bethe--Salpeter amplitudes (BSAs) and their adjoints are Fourier--transformed. In the two particle (and of course anti--particle) sector this is done e.g.\ by:
\begin{eqnarray}
\ll 0| \, T \, [ \, \psi^{(2)}  (-\,\eta_1\, x) \; \psi^{(1)}  (\eta_2\, x) \, ] \, |P,B>
& =: & \int\frac{d^4q}{(2\pi )^4} \;\; e^{\displaystyle -iqx} \;\,
\psi^{\,(\,2\,1\,)}_B (P,q) \nonumber \\
\ll 0| \, T \, [ \, \bar{\psi}^{(2)}  (-\,\eta_1\, x) \; \bar{\psi}^{(1)}  (\eta_2\, x) \, ] \, |P,B>
& =: & \int\frac{d^4q}{(2\pi )^4} \;\; e^{\displaystyle -iqx} \;\,
\psi^{\,(\,\bar{2}\,\bar{1}\,)}_B (P,q) \nonumber \\
 & \ldots &  \nonumber \\
\ll P,B| \, T \, [ \, \bar{\psi}^{(1)}  (\eta_2\, x) \; \bar{\psi}^{(2)}  (-\,\eta_1\, x) \, ] \, |0>
& =: & \int\frac{d^4q}{(2\pi )^4} \;\; e^{\displaystyle \,+iqx} \;\,
\tilde{\psi}^{\,(\,2\,1\,)}_B (P,q) \nonumber \\
  & \ldots &  
\end{eqnarray}
Treating the deuteron as a pure two nucleon system the ABSSM yields e.g.\ ($M=0,\pm 1$):
\begin{eqnarray} |d^+(P,M)> & \simeq &
\sum\limits_{s_1,t_1} \sum\limits_{s_2,t_2}
\int 
\frac{d^3q}{(2\pi )^3} \; \frac{1}{2\, m_{{}_N} \, 2 \,m_{{}_N}} \; b^+ \, (\vec{p}_1,s_1,t_1) \; b^+ \, (\vec{p}_2,s_2,t_2) |0> \; \cdot
 \nonumber \\
 & \cdot &  
\bar{u}^{(1)} \, (\vec{p}_1,s_1,t_1) \;
\bar{u}^{(2)} \, (\vec{p}_2,s_2,t_2) \;
\phi^{\,(\,2\,1\,)}_{d^+\!,M} (P,\vec{q}\,) 
\end{eqnarray}
Here I used $\displaystyle\phi^{\,(\,i\,j\,)}_B (P,\vec{q}\,) := \int dq^0 \; \psi^{\,(\,i\,j\,)\,}_B (P,q\,) / (2\pi)$ ($i\in \{2,\bar{2}\}$, $j\in \{1,\bar{1}\}$). Application of (9) to the normalization condition (3) yields within the ABSSM an {\em interaction independent} normalization condition involving all appearing n--(anti--)particle BSAs.  For (11) I obtain: 
\begin{eqnarray} \lefteqn{2 \, \omega_{d^+} (\oabs{P}) \;\delta_{M^\prime M} \; \simeq \; 
\int 
\frac{d^3q}{(2\pi )^3} 
\; \frac{2 \, \omega_{{}_N} (|\vec{p}_1|)}{2 \,m_{{}_N}} 
\; \frac{2 \, \omega_{{}_N} (|\vec{p}_2|)}{2 \,m_{{}_N}} }
\nonumber \\
& & \Big\{ \,\;\;\;
\tilde{\phi}^{\,(\,2\,1\,)}_{d^+\!,M^\prime} (P,\vec{q} \,) \;\;
\, \left. \frac{(\slsh{p}_1 + m_{{}_N})^{(1)}}{2\,m_{{}_N}} \right|_{p^0_1=\omega_{{}_N} (|\vec{p}_1|)} \; 
\left. \frac{(\slsh{p}_2 + m_{{}_N})^{(2)}}{2\,m_{{}_N}} \right|_{p^0_2=\omega_{{}_N} (|\vec{p}_2|)} 
 \phi^{\,(\,2\,1\,)}_{d^+\!,M } (P,\vec{q}\,)  \; - \nonumber \\
& & - \,
\tilde{\phi}^{\,(\,2\,1\,)}_{d^+\!,M^\prime} (P,-\,\vec{q}\,)
\, \left. \frac{(\slsh{p}_1 + m_{{}_N})^{(21)}}{2\,m_{{}_N}} \right|_{p^0_1=\omega_{{}_N} (|\vec{p}_1|)} 
\left. \frac{(\slsh{p}_2 + m_{{}_N})^{(12)}}{2\,m_{{}_N}} \right|_{p^0_2=\omega_{{}_N} (|\vec{p}_2|)} 
 \phi^{\,(\,2\,1\,)}_{d^+\!,M} (P,\vec{q}\,)  
 \, \Big\} 
\end{eqnarray}
This should be compared to the normalization condition for translational invariant BSAs: 
\begin{eqnarray} \lefteqn{2 \, P^\mu \,\delta_{B^\prime B} \;\simeq }  \nonumber \\
 &  & - \,
\frac{2\, i}{(2\,\pi)^4} \int d^4q \;
\tilde{\psi}^{\,(\,2\,1\,)}_{B^\prime} (P,q)
\left[ \frac{\partial}{\partial P_{\mu}} \Big( (\slsh{p}_1 - m_1)^{(1)} 
 (\slsh{p}_2 - m_2)^{(2)} \Big) \, \right]\Bigg|_{P^2=M^2_B}
 \psi^{\,(\,2\,1\,)}_B (P,q) \quad
\end{eqnarray}
Taking only into account positive energy eigenstates the following ansatz can be made:
\begin{eqnarray}
\lefteqn{\phi^{\,(\,2\,1\,)}_{d^+\!,M} (P,\vec{q}\,) \; \simeq
\;
 \frac{i}{\sqrt{2}} \; 
 \sqrt{\frac{2 \, \omega_{d^+} (\oabs{P})}{2 \, \omega_{{}_N} (|\vec{p}_1|) 
\; 2 \, \omega_{{}_N} (|\vec{p}_2|)}} \; 
\frac{(\slsh{p}_1 + m_{{}_N} )^{\,(1)}}{\sqrt{m_{{}_N} + \omega_{{}_N} (|\vec{p}_1 |)}} \;
\frac{(\slsh{p}_2 + m_{{}_N} )^{\,(2)}}{\sqrt{m_{{}_N} + \omega_{{}_N} (|\vec{p}_2 |)}} 
} \nonumber \\
 & &  
\qquad\qquad\quad \;\cdot \; 4\pi \; 
 \left\{ \;
 u\, (|\vec{q}\,|,P) \; - \;
 w\, (|\vec{q}\,|,P) \;  \sqrt{\frac{1}{8}} \;
{\cal S}_{12} (\hat{\vec{q}}\,)
 \; \right\} \; |1M;00>  \\ 
\lefteqn{|1M;00> \; :=
 \sum\limits_{s_1,t_1} \sum\limits_{s_2,t_2}  <\frac{1}{2}\, s_1,\frac{1}{2}\, s_2\, |\, 1M > \,
 <\frac{1}{2}\, t_1,\frac{1}{2}\, t_2\, |\, 00 > \;
\frac{u^{(1)} \, (\vec{0},s_1,t_1)}{\sqrt{\,2\, m_{{}_N}}} \;
\frac{u^{(2)} \, (\vec{0},s_2,t_2)}{\sqrt{\,2\, m_{{}_N}}} 
} \nonumber 
\end{eqnarray} 
${\cal S}_{12} (\hat{\vec{q}}\,) \; := \; 
\left[ 3 \, \vec{\gamma}^{\,(1)} \cdot \vec{q} \;\, \vec{\gamma}^{\,(2)} \cdot \vec{q} -
\vec{\gamma}^{\,(1)} \cdot \vec{\gamma}^{\,(2)} \; |\vec{q}\,|^2 \right] \, \gamma_5^{(1)} \, \gamma_5^{(2)}  / |\vec{q}\,|^2$ is the 4-dim.\ analogue of the tensor operator. Application of (14) to (12) yields the standard normalization condition $1=\int^\infty_0 d|\vec{q}\,| \;\, 8\, |\vec{q}\,|^2 \,(|u|^2+|w|^2)$ for the S-wave $u\, (|\vec{q}\,|,P)$ and the D-wave $w\, (|\vec{q}\,|,P)$. By the use of tensor spherical harmonics a direct connection between $u\, (|\vec{q}\,|,P)$, $w\, (|\vec{q}\,|,P)$ and the standard BS (and various Quasi-potential (QP)) vertex functions could be established: 
\beqa u\, (\oabs{q},P)  & = & - \, i \,
 \sqrt{\frac{1}{(4\pi)^{\, 3} \, \omega_{d^+} (\oabs{P})}} \, 
\int \frac{dq^0}{2\pi} \;
\frac{\Gamma^{\,(\,2\,1\,)}_{d^+\,+ +} (P,q^0,|\vec{q}\,|,M; {}^{3}{S}_{\,1}\,) 
}{(p_{\,2}^0 - \omega_{{}_N}(|{\vec{p}}_{\,2}|)+i\,\varepsilon) \; (p_{\,1}^0 - \omega_{{}_N}(|{\vec{p}}_{\,1}|)+i\,\varepsilon)} \nonumber \\ 
 w\, (\oabs{q},P)  & = & +\,i\, 
 \sqrt{\frac{1}{(4\pi)^{\, 3} \, \omega_{d^+} (\oabs{P})}} \, 
\int \frac{dq^0}{2\pi} \;
\frac{\Gamma^{\,(\,2\,1\,)}_{d^+\,+ +} (P,q^0,|\vec{q}\,|,M; {}^{3}{D}_{\,1}\,) 
}{(p_{\,2}^0 - \omega_{{}_N}(|{\vec{p}}_{\,2}|)+i\,\varepsilon) \; (p_{\,1}^0 - \omega_{{}_N}(|{\vec{p}}_{\,1}|)+i\,\varepsilon)} \nonumber  
\eeqa
Applying the ABSSM to (2) and (1) the electromagnetic formfactors of the deuteron in Impulse Approximation and the total cross section of $NN\rightarrow d\,\pi,\eta,\eta^\prime,\ldots$ were calculated using various BS and QP deuteron amplitudes. Results will be discussed in more detail in a forthcoming publication. As first conclusive remarks it should be stated, that the normalization conditions (12) and (13) can be fulfilled to high precision simultaneously. The results of the ABSSM are quite comparable to 4--dim.\ BS results and previous QP calculations. Due to the Wigner--rotation between in-- and out--states in (2) interference terms between the S-- and D--wave appear even in the charge formfactor of the deuteron, while such interference terms generate additional ``intrinsic'' contributions to its magnetic and quadrupole moment. The results are valid to all orders in the momentum transfer.
\end{document}